\documentclass[article,aps,amssymb,amsmath,pra,twocolumn,showpacs,superscriptaddress,longbibliography]{revtex4-2}

\usepackage{ulem}
\usepackage{comment} % Add this line to your preamble

\usepackage{graphicx}% Include figure files
\graphicspath{{figs_main/}{figs_SI/}}

\usepackage{dcolumn}% Align table columns on decimal point
\usepackage{bm}% bold math
\usepackage{subfigure}
\usepackage{color}
\usepackage{soul}

\usepackage{amsfonts,verbatim}

\usepackage[margin=0.6in]{geometry}

\usepackage{tikz}

\usepackage[english]{babel}
\usepackage{dsfont}
\usepackage{latexsym}
\usepackage{float}
\usepackage{afterpage}
\usepackage{enumitem}
\usepackage{mleftright}

\definecolor{lR}{rgb}{1, 0.8, 0.79}

\usepackage{eso-pic, graphicx}

\usepackage{listings}
\usepackage{multirow}
\usepackage{xcolor,colortbl}

\usepackage{bbm}
\usepackage{upgreek}
\usepackage{newtxtext,newtxmath}

\newcommand{\ignore}[1]{}

\newcommand{\nocontentsline}[3]{}
\newcommand{\tocless}[2]{\bgroup\let\addcontentsline=\nocontentsline#1{#2}\egroup}

\definecolor{Ablue}{rgb}{0.96,0.24,0.00}

\definecolor{Abluetitle}{rgb}{0.,0.24,0.51}

\definecolor{orange}{rgb}{0.96,0.24,0.00}

\definecolor{darkred}{rgb}{0.55, 0.0, 0.0}

\definecolor{darksalmon}{rgb}{0.91, 0.59, 0.48}
\definecolor{maroon}{cmyk}{0,0.87,0.68,0.32}

\setcitestyle{numbers,square,citesep={,\kern-.24em}}

\definecolor{mustard}{rgb}{1.0, 0.86, 0.35}

\addto\captionsenglish{\renewcommand{\figurename}{Fig.}}

\definecolor{Gray}{gray}{0.85}
\definecolor{LightCyan}{rgb}{0.88,1,1}
\newcolumntype{a}{$>${\columncolor{Gray}}c}
\newcolumntype{b}{$>${\columncolor{White}}c}

\usepackage{array}
\newcolumntype{L}[1]{$>${\raggedright\let\newline\\\arraybackslash\hspace{0pt}}m{#1}}
\newcolumntype{C}[1]{$>${\centering\let\newline\\\arraybackslash\hspace{0pt}}m{#1}}
\newcolumntype{R}[1]{$>${\raggedleft\let\newline\\\arraybackslash\hspace{0pt}}m{#1}}

\newcolumntype{P}[1]{>{\centering\arraybackslash}p{#1}}
\newcolumntype{M}[1]{>{\centering\arraybackslash}m{#1}}

\usepackage[colorlinks=true , citecolor=black,urlcolor=blue]{hyperref}

\newcommand{\tm}{{\text -}}

\newcommand{\tacq}{t_{\R{acq}}}

\newcommand{\app}{\approx}

\newcommand{\tpol}{t_{\R{pol}}}

\newcommand{\Cs}{{}^{13}\R{C}}

\newcommand{\Hs}{{}^{1}\R{H}}

\newcommand{\sq}[1]{\sqrt{#1}}

\newcommand{\beq}{\begin{equation}}
\newcommand{\eeq}{\end{equation}}
                  
\newcommand{\benum}{\begin{enumerate}}
\newcommand{\eenum}{\end{enumerate}}
                    
\newcommand{\bit}{\begin{itemize}}
\newcommand{\eit}{\end{itemize}}
\newcommand{\xhat}{\hat{\T{x}}}
\newcommand{\yhat}{\hat{\T{y}}}

\newcommand{\bea}{\begin{eqnarray}}
\newcommand{\eea}{\end{eqnarray}}

\newcommand{\bev}{\begin{verbatim}}
\newcommand{\eev}{\end{verbatim}}

\newcommand{\T}[1]{\textbf{#1}}
\newcommand{\I}[1]{\textit{#1}}
\newcommand{\R}[1]{\textrm{#1}}
\newcommand{\zfl}[1]{\protect\label{fig:#1}}
\newcommand{\zfr}[1]{\figurename\,\ref{fig:#1}}

\newcommand{\ba}{\left\{ \begin{array}{lr}}
\newcommand{\ea}{\end{array}\right.}

\newcommand{\blist}[1]{
 \begin{list}{#1}%$\ast\circ\bullet\Right
 \begin{align}
	 arrow
 \end{align}
 $\checkmark\star
  { \setlength{\itemsep}{3pt}
     \setlength{\parsep}{2pt}
     \setlength{\topsep}{3pt}
     \setlength{\partopsep}{0pt}
     \setlength{\leftmargin}{1em}
     \setlength{\labelwidth}{1em}
     \setlength{\labelsep}{0.5em} } }
\newcommand{\elist}{
  \end{list}  }

\DeclareMathSymbol{\vartheta}{\mathalpha}{letters}{"12}
\DeclareMathSymbol{\theta}{\mathalpha}{letters}{"23}
\DeclareMathSymbol{\phi}{\mathalpha}{letters}{"27}
\DeclareMathSymbol{\varphi}{\mathalpha}{letters}{"1E}

\newcommand{\bef}
{
\begin{figure}[htbp]
\centering
}

\newcommand{\eef}{\end{figure}}

\mleftright
\makeatletter
\@addtoreset{section}{part}
\makeatother

\newcommand{\affA}{Department of Chemistry, University of California, Berkeley, Berkeley, CA 94720, USA.}
\newcommand{\affB}{Hamamatsu Photonics, Sunayama-cho, Chuo-ku, Hamamatsu City, Shizuoka Pref., 430-8587, Japan.}
\newcommand{\affC}{ICFO– Institut de Ciències Fotòniques, The Barcelona Institute of Science and Technology, 08860 Castelldefels (Barcelona), Spain}
\newcommand{\affE}{Chemical Sciences Division,  Lawrence Berkeley National Laboratory,  Berkeley, CA 94720, USA.}
\newcommand{\affF}{CIFAR Azrieli Global Scholars Program, 661 University Ave, Toronto, ON M5G 1M1, Canada.}

\begin{document}
%\title{Zero-field NMR in natural isotopic abundance with multichannel modality}
\title{High-sensitivity multichannel zero-to-ultralow field NMR with atomic magnetometer arrays}
\author{Blake Andrews}\thanks{These authors contributed equally to this work}\affiliation{\affA}
\author{Matthew Lai}\thanks{These authors contributed equally to this work}\affiliation{\affA}
\author{Zhen Wang}\affiliation{\affA}
\author{Norihisa Kato}\affiliation{\affB}
\author{Michael Tayler}\affiliation{\affC}
\author{Emanuel Druga}\affiliation{\affA}
\author{Ashok Ajoy}\email{ashokaj@berkeley.edu}\affiliation{\affA}\affiliation{\affE}\affiliation{\affF}
 	 
\begin{abstract}
Despite its versatility and high chemical specificity, conventional NMR spectroscopy is limited in measurement throughput due to the need for high-homogeneity magnetic fields, necessitating sequential sample analysis, and bulky devices. Here, we propose a multichannel NMR device that overcomes these limitations that leverages the zero-to-ultralow field (ZULF) regime, where simultaneous detection of multiple samples is carried out via an array of compact optically pumped magnetometers (OPMs).
A magnetic field is used only for pre-polarization, permitting the use of large-bore, high-field, \I{inhomogeneous} magnets that can accommodate many samples concurrently. Through systematic advances, we demonstrate high-sensitivity, high-resolution ZULF NMR spectroscopy with sensitivity comparable to benchtop NMR systems. The spectroscopy remains robust without the need for field shimming for periods on the order of weeks. We show the detection of ZULF NMR signals from organic molecules without isotopic enrichment, and demonstrate the parallelized detection of three distinct samples simultaneously as a proof-of-concept, with the potential to scale further to over 100 channels at a cost comparable to high-resolution liquid state NMR systems. This work sets the stage for using multichannel ``NMR camera" devices for inline reaction monitoring, robotic chemistry, quality control, and high-throughput assays.
\end{abstract}

\maketitle
\pagebreak

\section{Introduction}

Nuclear Magnetic Resonance (NMR) spectroscopy is a key tool in chemical analysis due to its ability to examine chemical kinetics and molecular bonding non-invasively, and with high specificity \cite{slichterPrinciplesMagneticResonance1990, ernstPrinciplesNuclearMagnetic1990}. However, conventional NMR spectroscopy is expensive and limited in chemical throughput, primarily due to the need for highly homogeneous superconducting magnets. The stringent requirements for magnetic field homogeneity (typically at \textless 10ppb level \cite{shapiraSpatialEncodingAcquisition2004}) engender small ``sweet spots" (${<}$5cm$^3$) where samples must be placed sequentially for analysis. Additionally, the need for spatiotemporal homogeneity makes the magnets bulky, and necessitates involved strategies for shimming and deuterium field lock \cite{vanzijlUseDeuteriumNucleus1987}. These limitations are particularly problematic in emerging fields like robotic chemical synthesis and combinatorial screening ~\cite{macleodSelfdrivingLaboratoryAccelerated2020, burgerMobileRoboticChemist2020, haseNextGenerationExperimentationSelfDriving2019}, where there is a critical need for in-line monitoring tools to support continuous AI-driven reaction optimization, but for which NMR is currently less well-suited.

In this work, we aim to break the throughput limit by proposing an alternative strategy for parallelized NMR spectroscopy that exploits the zero-to-ultra-low field (ZULF) regime. Here, NMR detection is carried out in a magnetically shielded environment, while a magnet is solely used for pre-polarization. As a result, the magnetic field homogeneity requirements are completely relaxed. The ZULF NMR spectra are dominated by scalar J-couplings, and for one-dimensional spectra of small molecular systems, the spectra can be as informative and chemically specific as traditional high-field (HF) NMR \cite{putZeroUltralowFieldNMR2021,blanchardHighResolutionZeroFieldNMR2013}. Detection is carried out by compact, commercially available optically pumped magnetometer (OPM) devices \cite{blanchardZeroUltralowfieldNuclear2020}, and the lifting of homogeneity requirements means that one can arrange for an \I{array} of detectors to discern NMR spectra from multiple samples simultaneously. Additionally, the magnets used for sample pre-polarization can be \I{inhomogeneous}, and accommodate a large number of samples concurrently. We demonstrate the first proof of concept of this vision (Fig. 1). 

A significant barrier for ZULF NMR has been its relatively low sensitivity—often at least two orders of magnitude worse than benchtop NMR spectroscopy — diminishing its appeal despite its advantages. Here, through a series of technical improvements, we demonstrate the ability to address this challenge, for the first time, showing high signal-to-noise ratio (SNR) spectroscopy matching the $\Cs$ NMR sensitivity of a commerical 1.9T system. Our approach exploits pre-polarization with a large-bore (inhomogeneous) superconducting magnet, while implementing several methods to combat experimental noise. Notably, our approach supports high-resolution spectroscopy with the complete absence of shimming or OPM re-calibration for extended periods on the order of weeks.

Leveraging these advances, we report the first ZULF NMR spectra of organic molecules at natural $\Cs$ isotopic abundance without hyperpolarization.  We demonstrate parallelized detection of three distinct samples simultaneously as a proof-of-concept; however, we estimate the ability to scale to ${>}$100 channels with comparable cost to a high-resolution liquid state NMR system (see SI Sec. II \cite{SI}).

\begin{figure*}[t]
  \centering
  {\includegraphics[width=0.97\textwidth]{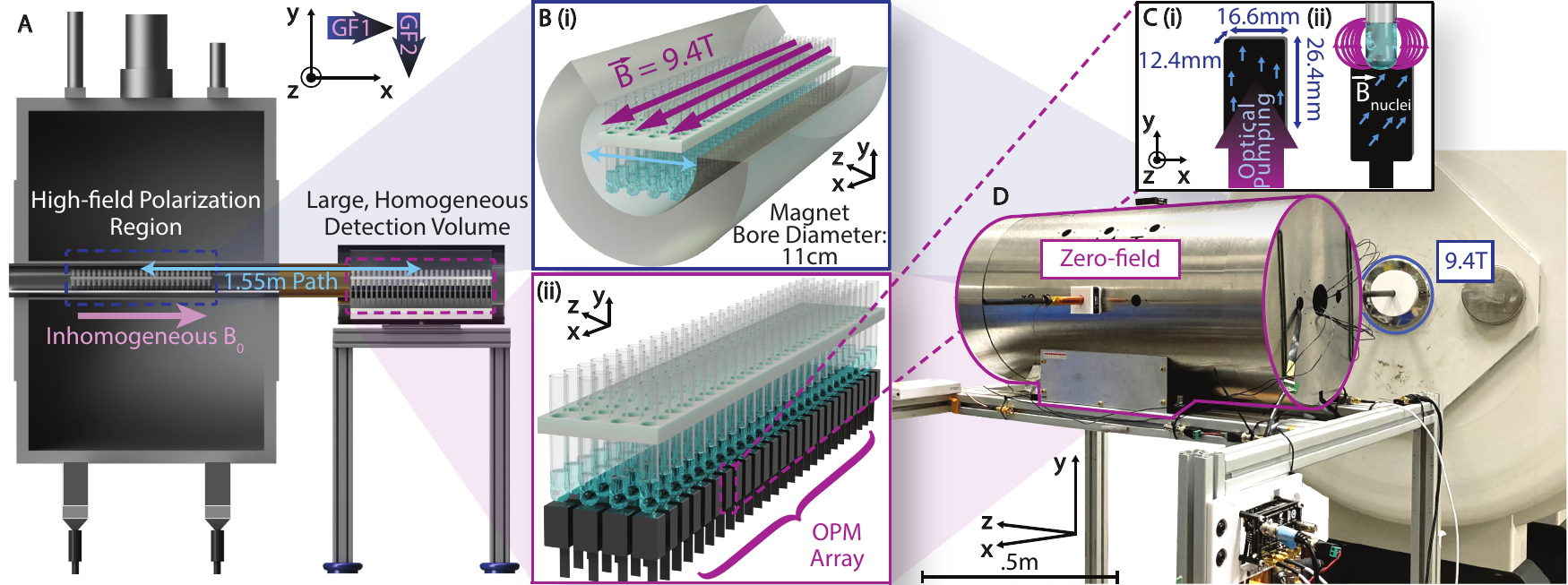}}
  \caption{\T{High-sensitivity, multichannel, ZULF NMR system.} (A) \I{Concept}. Instrument consists of a high-field, inhomogeneous, pre-polarization magnet here at $B_0{=}9.4$T, along $\xhat$. The magnet is shielded, horizontally oriented, and hosts a large bore (11cm diameter). ZF NMR detection is carried out with an array of OPMs in a mu-metal shielded region. Samples are mechanically shuttled between HF prepolarization and ZF detection centers, separated by a ${\app}$1.55m travel distance, in ${<}$1s. (B) \I{Schematic} of sample placement in (i) HF center for prepolarization, where magnet inhomogenity plays insignificant role; and (ii) ZF center, where samples are probed non-invasively by an array of OPMs. We estimate ${>}$100 samples can be accommodated at the HF and ZF regions (see SI Sec. II~\cite{SI}). (C) \I{OPM operation} for ZF NMR detection is schematically represented (see Ref \cite{budkerOpticalMagnetometry2007}). OPMs are compact (dimensions marked), allowing for arrayed operation (see \zfr{mfig5}). (D) \I{Photograph} of assembled instrument. See SI Sec. I ~\cite{SI} for discussion on device construction and additional photographs.  Coordinate axes (marked) highlight connection between the different panels.
}
\zfl{mfig1}
\end{figure*}

\section{Experimental Design}
\zfr{mfig1}A shows a conceptual depiction of the multichannel NMR device, employing an inexpensive, inhomogeneous, $B_0{=}9.4$T horizontal-bore magnet for pre-polarization, positioned adjacent to a mu-metal magnetic shield for NMR measurements in the ZULF regime, or alternately, the zero-field (ZF) regime. The superconducting magnet enables higher pre-polarization fields compared to previous experiments in the literature, enhancing ZF NMR signal strength. Inner shielding in the magnet allows a small separation between the HF and ZF centers (${\sim}$1.55m center-to-center) despite heightened fringe fields in the horizontal configuration. In operation, the samples are “shuttled” between both centers by a high-speed motor driven belt actuator in \textless 1s; in practice, a solenoid “guiding field” (GF$_1$) provides a switchable weak field over the path adjoining the centers (see \zfr{mfig2}A, SI Sec. I \cite{SI}). 
Since detection occurs at the ZF center, magnet inhomogeneity manifests as negligible variations in overall ZF NMR signal intensity without affecting spectral resolution, effectively subverting the otherwise usual line-broadening effects stemming from inhomogeneity. Hence, the apparatus can tolerate magnets with inhomogeneities as large as a few percent. This permits using cost-effective high-field magnets that host large-bores. In our device, for instance, the magnet can accommodate a rack of \textgreater 100 sample NMR tubes simultaneously; this concept is schematically depicted in \zfr{mfig1}B (i).

Since homogeneity considerations are comparably trivial to handle at zero-field, the ZF chamber can be made spacious enough to accommodate a similarly large number of samples at low cost (\zfr{mfig1}A). Tunable low fields can can be created within it; allowing us to access both the ZF and ultra-low-field (ULF) regimes \cite{bodenstedtFastfieldcyclingUltralowfieldNuclear2021}. Key to multichannel operation in our work is conducting NMR measurements through an array of commercially available OPMs (QuSpin \cite{shahQZFMGen32062021}), schematically shown in \zfr{mfig1}B (ii) as a concept, placed in proximity to the intially prepolarized NMR tube rack. Such arrayed detection exploits the compact form factor of the OPMs (\zfr{mfig1}C(i)), which can detect NMR signals at a relatively high standoff (${>}$5mm).

For clarity, \zfr{mfig1}C shows a schematic operation of one of the OPMs from the array \cite{tierneyOpticallyPumpedMagnetometers2019}. An integrated laser optically polarizes Rb atoms in a vapor cell within the unit as seen in \zfr{mfig1}C (i). Transmission of the light through the vapor cell is measured in the presence of a weak, localized oscillating transverse magnetic field (\textless 100 nT, ${\sim}$1kHz). The quadrature demodulation signal is proportional to the background field component along the modulation axis, allowing sensitive measurement of fields from the ensemble of analyte nuclei in the sample container. Alternatively, one can envision configurations using a single cell excited by an array of laser beams instead of discrete OPM units \cite{kimParallelHighfrequencyMagnetic2020}. 

\zfr{mfig1}D shows a photograph of the assembled device implementing a proof-of-concept of the vision in \zfr{mfig1}A. Pictured is the horizontal magnet and ZF center (see also \zfr{mfig5}B). More details of device construction and photographs are provided in SI Sec. I ~\cite{SI}. 

\begin{figure*}[t]
  \centering
  {\includegraphics[width=0.97\textwidth]{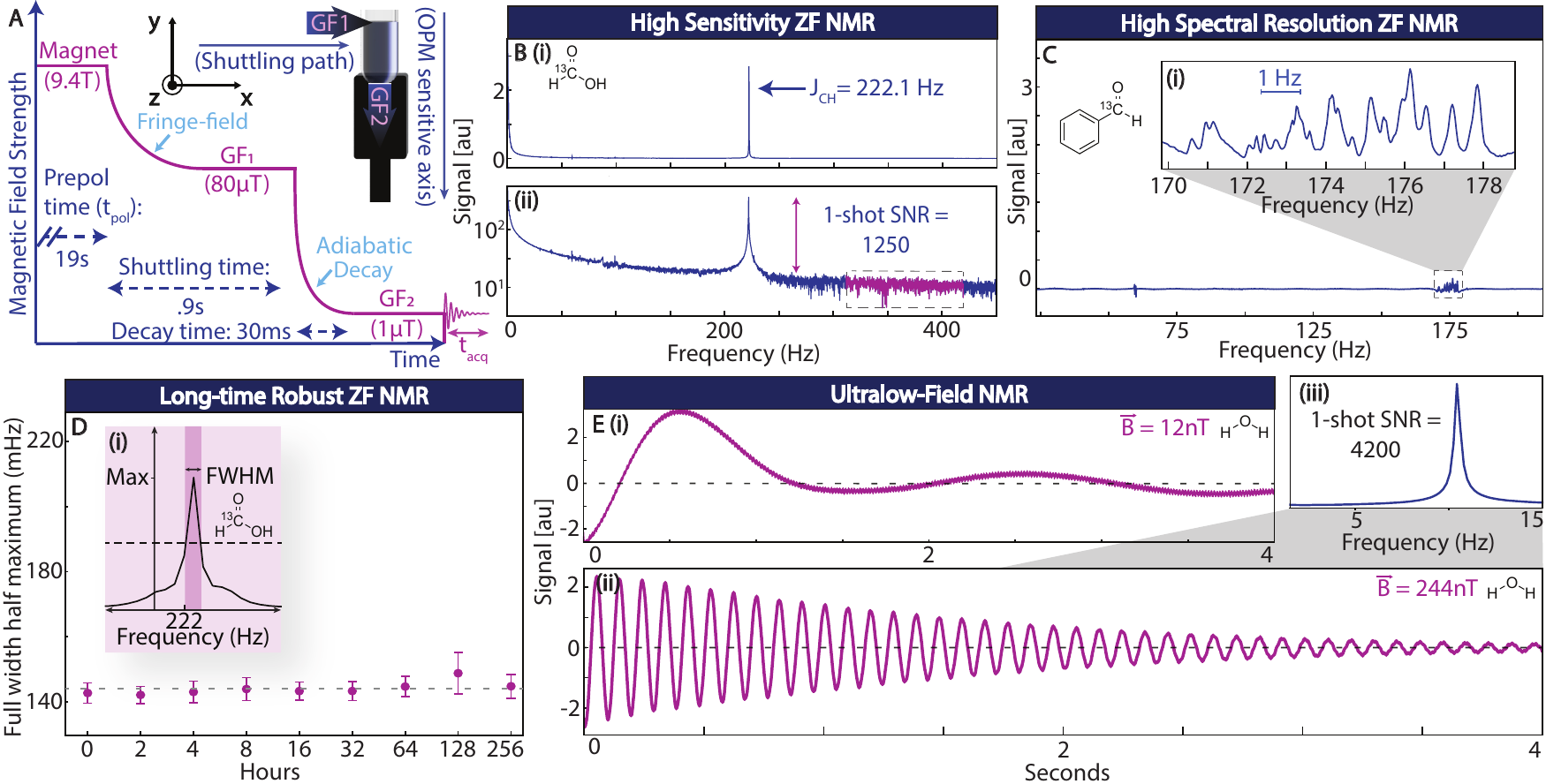}}
  \caption{\T{Robust, high SNR, ZF NMR via inhomogenous field polarization.} (A) \I{Experiment schematic}. Sample is prepolarized in the inhomogenous $B_0{=}9.4$T field for $\tpol$ (19s for formic acid) before moving through the first guiding field (GF$_1$) solenoid (80$\upmu$T) along the shuttling path for the duration of travel in (${\app}0.9$s). Before GF$_1$ is adiabatically turned off over 30ms, a 1$\upmu$T second guiding field (GF$_2$) is preemptively turned on uniformly throughout the shield. When the sample arrives, GF$_2$ is suddenly turned off immediately before acquisition, which proceeds for $\tacq{\sim} T_2^*$. (B) \I{High SNR spectroscopy}. Panel shows ZF-NMR spectrum of enriched formic acid over 20 scans with $\tacq{=}11$s in (i) linear and (ii) logarithmic scales. Characteristic peak at $J_{\R{CH}}{=}222.1$Hz is visible. Single shot SNR${=}1250$, calculated from the spectral wing (shaded). (C) \I{High-resolution ZF NMR} of enriched benzaldehyde with $\tacq{=}30$s shown over 200 scans. Phase correction and baseline subtraction is applied. \I{Inset (i):} Zoom in shows sub-hertz, $T_2^*$ limited linewidths. (D) \I{Long-time robust ZF NMR}. Spectral FWHM of enriched formic acid (inset) is used as a reporter of temporal stability over 256hrs (${\app}10.67$days). No shimming, calibration, or field compensation is applied during the entire period. Linewidths ($T_2^*$ limited) remain stable at 0.144Hz to within error, and are reported as an average of 128 scans (dashed line). Error bar: linewidth standard deviation over 128 scans. (E) \I{Ultra-low-field NMR} here conducted on water at (i) 12nT and (ii) 244nT. Panel shows single-shot free induction decays with (iii) high one-shot SNR value ${=} 4200$ at 244nT. All samples were prepared and measured under ambient atmospheric conditions (no deoxygenation) in a standard 5mm NMR tubes.
}
\zfl{mfig2}
\end{figure*}

\begin{figure}[t]
  \centering
  {\includegraphics[width=0.49\textwidth]{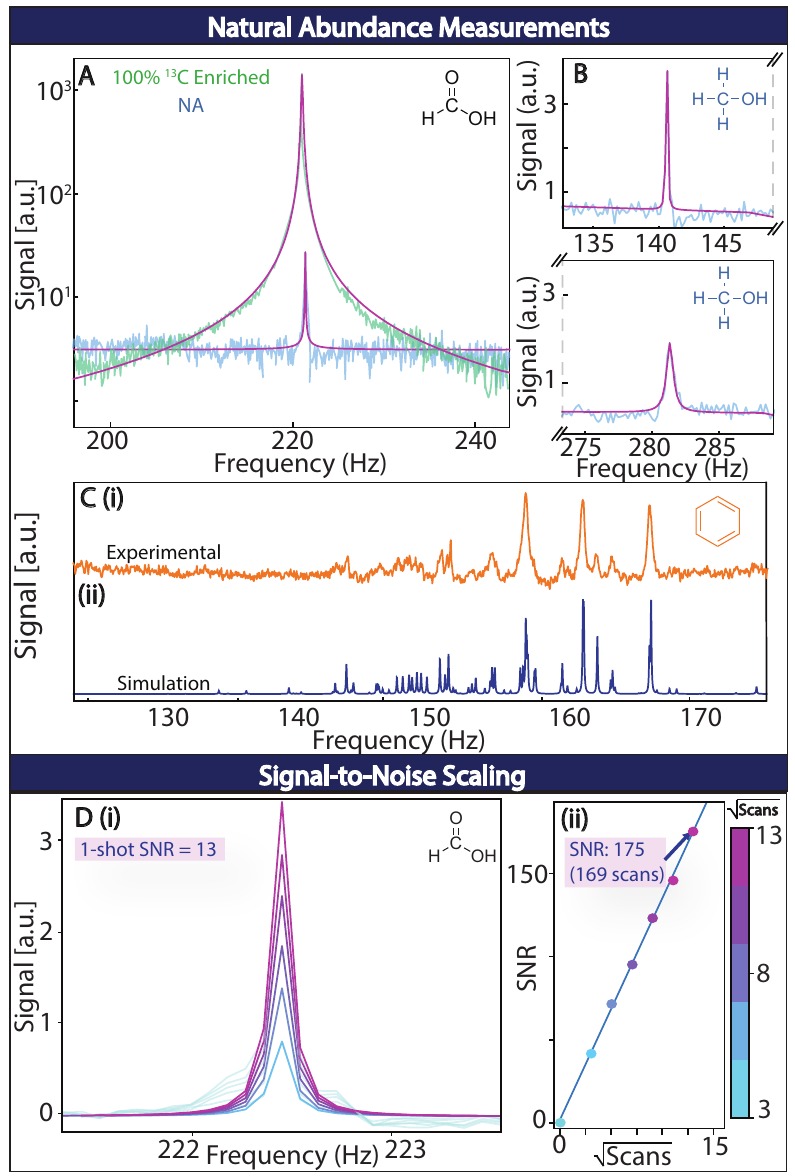}}
  \caption{\T{ZF NMR at natural $\Cs$ abundance in organic molecules.} (A) \I{SNR comparison}. Spectra corresponding to $\Cs$ enriched and natural abundance (NA) (1.1\%$\Cs$) (turquoise, blue respectively) of formic acid plotted on a logarithmic scale (10 scans each). NA signal is 100 times weaker, but discernible. (B) ZF spectrum of NA methanol (497 scans) in a neat solution shown in two windows centered at 140Hz and 280Hz. Lorentzian fits for formic acid and methanol are in purple. (C) ZF spectrum of NA benzene in a neat solution, taken over 1436 scans. (i) Experimental in orange with (ii) simulated data below in dark blue. (D) (i) NA formic acid ZF spectra taken with varying number of scans $N$ and (ii) SNR plotted against $\sq{N}$. SNR ${=}175$ after 169 scans.}
\zfl{mfig3}
\end{figure}

\begin{figure}[t]
  \centering
  {\includegraphics[width=0.49\textwidth]{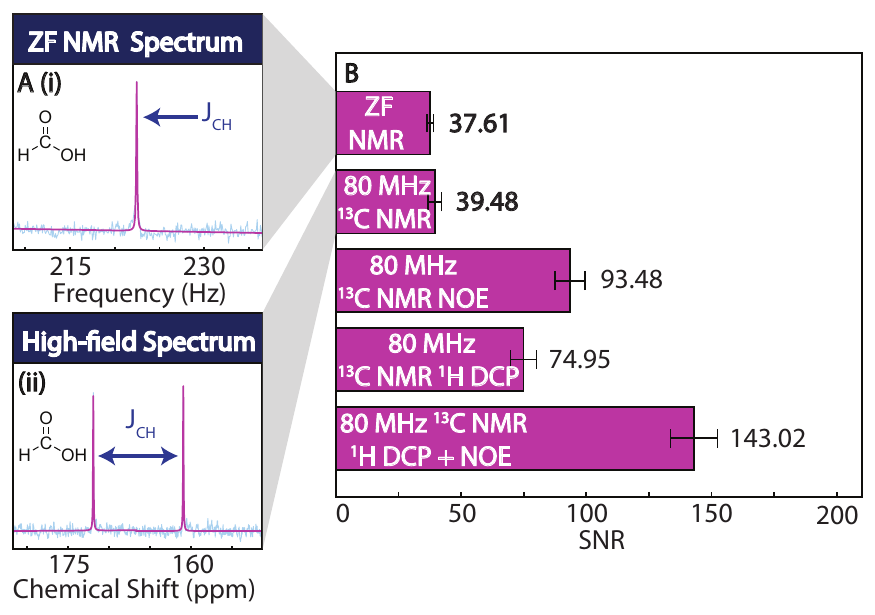}}
  \caption{\T{SNR comparison with benchtop HF NMR.} (A) \I{ZF and HF NMR spectrum} of NA formic acid, showing complementary features of J-couplings and chemical shifts. (B) \I{SNR comparison} between our apparatus and various $^{13}$C NMR experiments done on a Spinsolve 80 Carbon (${\sim}1.9$T detection field). Top to bottom: ZF-NMR via our apparatus, $^{13}$C NMR, $^{13}$C NMR with NOE enhancement, $^{13}$C NMR with proton decoupling, and $^{13}$C with NOE enhancement and proton decoupling. SNR is reported over 8 scans, while error bars are calculated over 30 repetitions of such experiments.}  
\zfl{mfig4}
\end{figure}

\begin{figure*}[t]
  \centering
  {\includegraphics[width=0.98\textwidth]{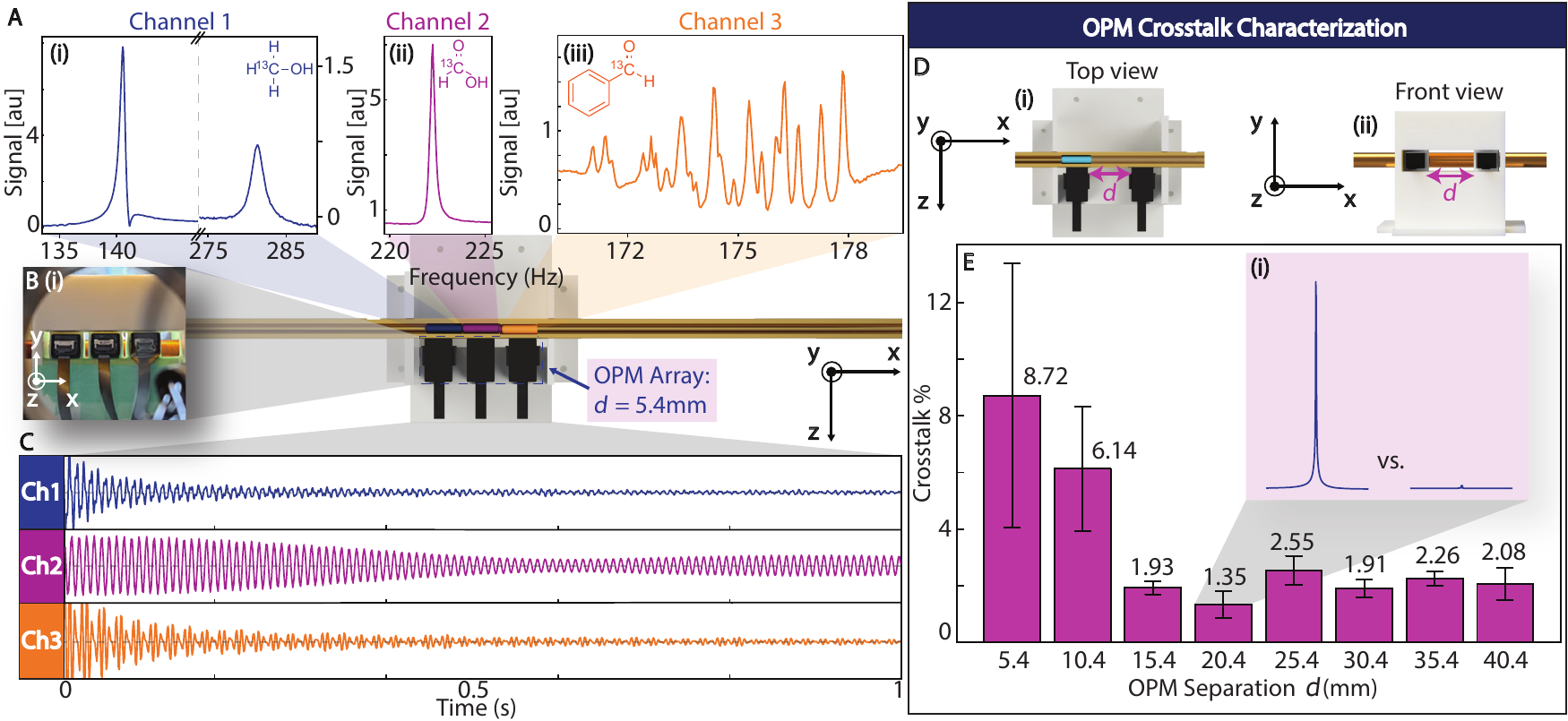}}
  \caption{\T{Multichannel ZF NMR via inhomogenous field prepolarization.} (A) \I{Simultaneous ZF NMR} spectra taken from three separate OPM channels, showing measurement of three distinct $^{13}$C enriched chemical samples: (i) methanol (dark blue), (ii) formic acid (purple), and (iii) benzaldehyde (orange). (B) \I{Inset (i):} \I{Photograph} of multichannel setup showing sideview of OPM arrangement. \I{Main panel}: Schematic of top-view of OPM arrangement, arranged in a linear fashion, separated by 5.4mm. Three samples are color-coded. Coordinate axes, corresponding to \zfr{mfig1} shown for clarity. (C) \I{Simultaneously acquired FIDs} of the three samples. Digital bandpass filters employed (top to bottom): 120-300Hz, 221-225Hz, 150-200Hz. (D) \I{Measurement of OPM crosstalk}. (i-ii) Pictorial representation of OPM separation $d$ defining the edge-to-edge distance between sensors. Sample of $^{13}$C enriched formic acid is placed in front of one OPM; cross-talk quantifies extent of signal sensed by the adjacent OPM. (E) \I{OPM crosstalk} calculated as ratio of baseline corrected peak heights expressed as a percentage plotted against separation for two adjacent OPMs. \I{Inset (i):} example spectra measured by the OPM directly front of sample (left) and the one adjacent to it (right). Here $d{=}20.4$mm.}
\zfl{mfig5}
\end{figure*}

\zfr{mfig2}A displays a schematic of the experimental sequence. Coordinate axes show orientations of the applied fields, OPM and sample for clarity (see also \zfr{mfig1}). Samples are initially polarized using the $B_0{=}9.4$T magnet for $\tpol$=3-5 times $T_1$ (${\sim}$19s for formic acid), followed by transfer through the magnet fringe field along the GF$_1$ solenoid energized at 80$\upmu$T, collinear with $B_0$ (here $\xhat$). Adiabatic turn-off of this field then transitions the sample to a secondary guiding field (GF$_2$) within the ZF center at 1$\upmu$T, that is preemptively activated to reorient nuclear spins along the OPM sensitive axis (here -$\yhat$). GF$_2$ is subsequently rapidly turned off, prompting a nonadiabatic transition to ZF and initiating spin dynamics that generate the ZF NMR signal, captured by the OPM. Inset in \zfr{mfig2}A details the orientations GF$_{1,2}$ relative to the OPM and sample. The entire field-switching protocol is driven using a low-cost, compact, controller (NMRduino~\cite{taylerNMRduinoModularOpensource2024}), equipped with high-current driver chips (DRV8838 and TB6612FNG for analog and digital channels respectively) and a LTC1859 chip for signal acquisition.

While the sequence in \zfr{mfig2}A itself follows prior work \cite{blanchardZeroUltralowFieldNMR2016,taylerInvitedReviewArticle2017}, we include several innovations in the device to enable high-SNR multichannel operation (see SI Sec. I, III ~\cite{SI}). The shuttling stage and motor are placed at the opposite end of the $B_0{=}9.4$T magnet, distancing them from the ZF center to mitigate electronic noise and vibration. GF$_{1,2}$ coils, OPMs, and all control electronics operate on battery power to further reduce interference. Vibration is minimized by mounting the ZF center on Sorbothane vibration-dampening feet and using a flexible carbon fiber rod to support the samples during shuttling (see SI Sec. I ~\cite{SI}). Finally, we adopt a ``pulse-free" technique that does not require a DC pulse to initiate the free induction decay (FID). This removes the need for pulsing coils within the ZF center, simplifying multichannel operations (see \zfr{mfig5}) by applying GFs uniformly throughout the shuttling path and shield, effectively manipulating all chemical samples in an identical manner. This additionally decreases the risk of magnetizing the shield through strong pulses, ultimately leading to highly robust operation on the order of weeks (\zfr{mfig2}D). We found the removal of these pulsing coils also reduced background noise close to the sensor, contributing to the high SNR.

\zfr{mfig2}B illustrates the outcomes of these instrumentation advances, initially focusing on a single channel. Using $^{13}$C enriched formic acid, a common benchmark used for J-spectroscopy at ZF, we achieve a single-shot SNR of ${\sim}$1250, presented on both linear (\zfr{mfig2}B (i)) and logarthmic scales (\zfr{mfig2}B(ii)). The sole visible peak is at the $\Cs\tm\Hs$ J-coupling value of 222.1Hz. To our knowledge, this SNR is the highest reported in the literature and is achieved here for an off-the-shelf sample without sample deoxygenation \cite{putZeroUltralowFieldNMR2021,blanchardZeroUltralowfieldNuclear2020,blanchardHighResolutionZeroFieldNMR2013}. \zfr{mfig2}B(i) also shows the spectrum is also notably free from spectral contamination, including line and motor noise. \zfr{mfig2}C displays the spectrum from a more complex sample, singly labeled $^{13}$C benzaldehyde. The main panel displays a broad frequency range, once again emphasizing high SNR and spectral purity, which matches previous reports but with minimal scans and sample preparation. Inset \zfr{mfig2}C(i)) zooms into the relavant window near 175Hz, showing distinct J-spectral features \cite{blanchardHighResolutionZeroFieldNMR2013}, with a $T_2^*$ limited linewidth \textless 0.2Hz.

A feature of our instrument is its robust temporal stability, highlighted in \zfr{mfig2}D. Here continuous ZF NMR measurements on enriched $^{13}$C-formic acid reveal a stable, $T_2^*$ limited, spectral linewidth of ${\sim}$145mHz, calculated as the FWHM (inset shows representative spectrum). Importantly, this linewidth remains stable for ${>}$10 days \I{without} any shimming or degaussing of the mu-metal shield \cite{DegaussingMagneticShields}, nor OPM re-calibration during the entire period, suggesting the linewidth will remain stable for several weeks. This stability contrasts with typical benchtop NMR systems where field instability is around 0.0014ppm/hr (0.083Hz at 60MHz) assuming \textless1$^\circ$C change in temperature, often necessitating frequent shimming or active field compensation. From our experience, the stability observed in \zfr{mfig2}D is also considerably better than ZF NMR apparatuses constructed with permanent (Halbach) magnets and pulsing coils. We attribute this to the low spatial gradient of the fringe field of the superconducting $B_0{=}9.4$T magnet, which is effectively shielded by the mu-metal in the ZF center, as well as reliable shield integrity in the absence of strong pulses.

Stable shield conditions also enable high-SNR ultralow field (ULF) NMR measurements. \zfr{mfig2}E(i-ii) demonstrates this with water at bias fields of 12nT and 244nT respectively, applied along the GF$_1$ coil. Here the FID is sampled every $\sim$165$\upmu$s, and the spins exhibit a slow Larmor precession of 0.51Hz and 10.39Hz that can be clearly discerned. The ULF measurement here is of similarly high SNR (\zfr{mfig2}E (iii)); we estimate a one-shot SNR ${\app}$4200 for the 244nT bias field. \zfr{mfig2}E suggests applications towards ULF NMR with reintroducing controlled Zeeman fields to gain in chemical resolution \cite{appeltPathsWeakStrong2010} or in exploiting enhanced relaxation-dispersion at ultralow fields \cite{bodenstedtFastfieldcyclingUltralowfieldNuclear2021}.

\section{ZULF NMR at natural abundance} 
The high sensitivity of our apparatus enables ZF NMR signal measurements at natural isotopic abundance (NA) with no deoxygenation, previously unfeasible due to low sensitivity. Historically, ZF NMR required $^{13}$C enriched molecules \cite{THEIS2013160,WILZEWSKI201766} or those containing specific spin-1/2 heteronuclei like ${}^{19}$F or ${}^{31}$P \cite{alcicekZeroFieldNMRSpectroscopy2021,trahmsNMRVeryLow2010}. Here we present, to our knowledge, the first direct ZULF detection of organic molecules at natural abundance of $^{13}$C nuclei. This is demonstrated in \zfr{mfig3}A with formic acid, comparing ZF NMR signals from fully enriched and NA samples on a logarithmic scale, showing the expected 100-fold SNR reduction. Dark solid lines here are Lorentzian fits.  NA methanol and benzene are similarly analyzed in \zfr{mfig3}B and \zfr{mfig3}C(i) respectively. The benzene spectrum corresponds well with simulations (\zfr{mfig3}C(ii)) despite being taken from a previously opened stock solution where degradation has likely occurred, faithfully reproducing chemistry laboratory conditions.

Given the stability of our detection apparatus, SNR can be enhanced through averaging. \zfr{mfig3}D (i) displays individual spectral traces with increasing averages $N$, demonstrating an expected SNR scaling ${\propto}\sq{N}$. For $N{=}169$, we achieve an SNR of 175 for NA formic acid, with an average single-shot SNR of 13.

We emphasize that significant further SNR improvements are achievable. Our experiments utilize standard 5mm NMR tubes that are not optimal and under-utilize the OPM detection volume. Additionally, for the current shielding conditions, the commercial OPMs we employ have a sensitivity ${\app}$40fT/$\sqrt{\R{Hz}}$ \cite{shahQZFMGen32062021}. Emerging prototype OPMs, however, have reported sensitivities closer to 1fT/$\sqrt{\R{Hz}}$ \cite{kominisSubfemtoteslaMultichannelAtomic2003}. By using these more sensitive OPMs and optimizing the sample tube to aid detection, the signal could be improved by over an order of magnitude. In addition, noise could be reduced further (up to 4-fold) by using a ferrite core in the innermost layer of the magnetic shield. In parallel, we anticipate even further signal gains from recent advances in high temperature superconducting (HTS) magnet technology \cite{chenWatchsized12Tesla2023,leeConstructionTestResult2022,parkDesignOverviewMIT2021} which can enable higher, albeit inhomogeneous, polarizing fields.

Even at current sensitivity, however, our apparatus compares well with benchtop high-field NMR technology. Traditionally, ZULF NMR sensitivity has been significantly lower—often by more than two orders of magnitude—compared to benchtop NMR; however, we demonstrate here a promising step towards bridging this gap. In particular, we compare our system to $^{13}$C NMR conducted on a Magritek Spinsolve 80 Carbon (detection field ${\sim}$1.9T), using \I{identical} samples of neat NA formic acid solutions in conventional 5 mm NMR tubes. 

\zfr{mfig4}A(i-ii) first compares the resulting NMR spectra at ZF and HF (1.9T), displaying J-resolved and chemical-shift-resolved spectra respectively, and demonstrating that ZF NMR spectra offers comparable chemical information for small organic molecules. The bar chart in \zfr{mfig4}B analyzes the signals over 8 averages. From the upper two bars in \zfr{mfig4}B, corresponding to the data in \zfr{mfig4}A, it is evident that the that SNR of ZF NMR in our apparatus is comparable to that of a standard, $^{13}$C benchtop NMR measurement, although furthur improvement is necessary to match the sensitivity of a $^{1}$H NMR experiment. Notably, our ZF apparatus does not require shimming or re-calibration between experiments (see \zfr{mfig2}D), nor deuterium locking, unlike the benchtop HF instrumentation where repeated adjustments are necessary. Importantly, while the HF NMR device can accomodate only single samples, which have to be measured serially, our ZF NMR device can allow multichannel operation (see \zfr{mfig5}). 

The other bars in \zfr{mfig4}B depict enhancements to HF signals via employing the nuclear Overhauser effect (NOE), proton decoupling ($^{1}$H DCP), and their combinations. The ZF-NMR signal remains within an order of magnitude of these enhanced cases. We anticipate that applying variants of these enhancement schemes in our setup, such as INEPT at HF prior to shuttling, could yield similar signal gains.

\section{Simultaneously acquired Multichannel ZULF NMR}

Leveraging pre-polarization with the inhomogeneous magnet, we now move to multichannel operation, presented in \zfr{mfig5} as a proof-of-concept of the vision outlined in \zfr{mfig1}. As a representative building block of larger arrayed detection, we constructed a 3x1 OPM array at the ZF center capable of \I{simultaneously} acquiring ZF-NMR spectra from three distinct samples -- here, enriched methanol, formic acid, and benzaldehyde. \zfr{mfig5}A shows a zoomed view into the J-resolved spectra of these samples. \zfr{mfig5}B details the setup within the ZF shield. For simplicity, the samples are arranged linearly, separated by $d{=}5.4$mm, corresponding to OPMs placed side-by-side. Inset \zfr{mfig5}B(i) shows a side-view photograph of the OPM arrangement, while the main panel presents a schematic top view with axes marked, corresponding to same coordinate system in \zfr{mfig1} and \zfr{mfig2}A, although the OPMs are placed beside the sample instead of below for ease of access and adjustments. Here, samples are contained in short 5mm tubes (length ${\sim}$15mm) and subjected to the sequence in \zfr{mfig2}A simultaneously. \zfr{mfig5}C displays the FID signals from the three samples, highlighting the simultaneous measurement capacity.

An important metric for feasibility of arrayed detection is ``crosstalk", where spectra may bleed into adjacent channels due to correlations at short OPM separations $d$, caused by sensor interference or spurious detection of a sample’s magnetization by neighboring sensors. 
In \zfr{mfig5}D-E, we use a pair of OPMs to quantify crosstalk between adjacent channels. The setup is depicted in \zfr{mfig5}D (i-ii), with coordinate axes marked for reference, and crosstalk is measured as a function of $d$ (\zfr{mfig5}E). We use a $^{13}$C-enriched formic acid sample placed in front of one OPM, measure the ZF-NMR signal from both OPMs simultaneously, and quantify crosstalk as the ratio of spectral magnitudes. Inset \zfr{mfig5}E(i) illustrates this for OPMs separated by $d{=}20.4$mm. 

The bar chart in \zfr{mfig5}D, derived from 10 consecutive shots, show that even at the closest side-by-side distance ($d{=}5.4$mm, the closest distance enabled in our custom 3D printed OPM holder in \zfr{mfig5}B(i)), the crosstalk is ${<}10\%$. It declines sharply thereafter with increasing $d$, and stabilizes at ${\sim}2\%$ at $d{=}15.4$mm. The saturation value likely reflects inherent crosstalk within the data acquisition unit itself, since this steady-state crosstalk value was observed in the channels even when the empty (adjacent) channel was completely turned off.

While these results serve as a proof-of-concept, prospects for scaling to arrayed OPM detection at a larger scale appear promising. Even with the current shield and magnet arrangement, and separating the OPMs by $d{=}5.4$mm, we estimate the capacity to accommodate ${>}$100 samples simultaneously (See SI Sec. II ~\cite{SI}). The low GF values ($\upmu$T) employed further facilitate this, as homogeneous low field solenoids can be easily fabricated to accommodate large sample arrays. 

Finally, let us contrast our approach to alternate methods for high throughput NMR. One strategy involves parallel NMR detection employing numerous microcoil receivers and small sample volumes at HF ~\cite{trumbullIntegratingMicrofabricatedFluidic2000,fratilaSmallVolumeNuclearMagnetic2011}. However, the magnet ``sweet spot" remains constant, setting an upper bound on the number of samples that can be accommodated simultaneously. Smaller coils might increase capacity, but they complicate the design and reduce efficiency. Alternately, approaches for simultaneous multi-nuclear detection enhance the richness of spectral information~\cite{kupceExperimentsDirectDetection2019,kupceParallelReceiversSparse2011}, but they do not expand the total sample throughput. An emerging strategy involves magnetic resonance spectroscopy imaging (MRSI)~ \cite{faderlAcceleratedScreeningProtein2024}. However, this method faces challenges in the expense of imaging-grade magnets, a low per-sample filling factor, and slower measurements due to the indirect dimension of spectroscopy.

In contrast, arrayed OPM-based ZULF detection is designed for direct spectroscopy, can be optimized for filling factor per sample, and is inherently extensible since the mu-metal shield can be made large enough to accommodate an arbitrarily large number of samples while also offering relaxed requirements for shimming and calibration.

\section{Outlook}
In certain contexts, multichannel ZULF NMR may already offer advantages over HF NMR, particularly for high throughput applications.  SI Sec. II~\cite{SI} presents a thorough cost analysis of an arrayed ZULF NMR system. Using conservative per-channel estimates based on commercial list prices and factoring in economies of scale, our findings reveal a benign cost scaling, suggesting the feasibility of constructing a ZULF NMR device with ${>}$100 channels for a cost comparable to that of a conventional liquid-state HF NMR system operating at 400-500MHz (see Fig. 2 in SI~\cite{SI}). We note that similar arrayed OPM devices are already attracting interest in magnetoencephalography (MEG) applications~\cite{brookesMagnetoencephalographyOpticallyPumped2022}.

We emphasize, however, that HF NMR systems have broader capabilities, including the ability to analyze a wide range of samples and perform multi-dimensional NMR, and currently exhibit higher sensitivity. However, we argue that for high-throughput 1D spectroscopy applications like in robotic chemistry~\cite{bornemann-pfeifferStandardizationControlGrignard2021, sansSelfOptimizingSynthetic2015}, where there is an unmet need for an in-line, non-invasive,  analysis tool for kinetic measurements, ZULF NMR could offer distinct advantages. The horizontal bore configuration of our system (\zfr{mfig1}D) is particularly beneficial, allowing ZF NMR analysis to be performed diametrically opposite the chemical synthesis region where samples can be loaded and shuttled. 

Furthermore, we anticipate straightforward improvements to our apparatus that could enhance sensitivity by more than an order of magnitude without compromising multichannel capacity. This includes optimizing the filling-factor and integrating next-generation OPMs \cite{hongFemtoteslaAtomicMagnetometer2024}. Additionally, HTS magnets, free from cryoshim requirements, could be optimized to house numerous samples in a compact form-factor. The capacity of multichannel ZF NMR systems to operate for extended periods without shimming or temporal field-lock (\zfr{mfig2}D) presents strong advantages for certain applications. It also portends new strategies to exploit cross-correlations between neighboring channels to suppress measurement noise \cite{zhouAmbientInterferencesSuppressing2020, srinivasReducingNoiseFloor2020, sampietroSpectrumAnalyzerNoise1999}.

Ultimately, this work suggests the potential for an \I{``NMR camera"} capable of observing multiple samples simultaneously with a benign cost scaling. This development could lead to novel assays for quality control \cite{rodriguesNMRMethodsBeer2011, heymanNMRbasedMetabolomicsQuality2012}, and disease detection \cite{ouyangNMRbasedMetabolomicStudy2011, hernandez-baixauliDetectionEarlyDisease2020,vignoliHighThroughputMetabolomics2019}, as well as applications that exploit the ZF regime’s ability to enhance relaxation dispersion \cite{bodenstedtFastfieldcyclingUltralowfieldNuclear2021}, eliminate susceptibility broadening artifacts, and penetrate metal containers \cite{buruevaChemicalReactionMonitoring2020, taylerUltralowfieldNuclearMagnetic2019}, potentially enabling \I{in-operando} battery diagnostics \cite{huSensitiveMagnetometryReveals2020} at scale.

\subsection*{Acknowledgements}
We thank Prof. Evan Williams for access to the magnet used in this work, and acknowledge early contributions from Ruhee Nirodi, Nicolas Matthey and Chongwei Zhang and technical support of Sven Bodenstedt. We gratefully acknowledge discussions with Christian Bengs, Raffi Budakian, Michael Semmlinger and Kathyrn Pritchard, and funding from NSF PFI (2141083), NSF EAGER (2231634), NSF MRI (2320520), Hamamatsu Photonics (20201452), and the CIFAR Azrieli Foundation (GS23-013).

\vspace{-5mm}

\vspace{-1mm}

\end{document}